\documentclass[letterpaper,english,twocolumn,amsmath,amssymb,superscriptaddress]{revtex4}
\usepackage[T1]{fontenc}
\usepackage[utf8]{inputenc}
\usepackage{babel}
\usepackage{array}
\usepackage{graphicx}
\usepackage{tikz}
\usepackage{float}
\usetikzlibrary{shapes,arrows}

\makeatletter

\usepackage{amsmath}

\usepackage{url}

\begin{document}

\newcommand{\ppi}{p_i}
\newcommand{\mamo}{colloid}
\newcommand{\mamos}{colloids}
\newcommand{\tps}{5cm}
\newcommand{\tndv}{6.3cm}
\newcommand{\tndh}{2.8cm}
\newcommand{\qfour}{$\overline{q}_4$}
\newcommand{\qsix}{$\overline{q}_6$}

\tikzstyle{eblock} = [rectangle, text width=\tps, text centered, minimum height=0.96*\tps]
\tikzstyle{line} = [draw, -latex']

\title{Two-stage crystallization of charged \mamos~at low supersaturations}

\author{Kai Kratzer}
\affiliation{ICP, Institute for Computational Physics, University of Stuttgart, 
Allmandring 3, 70569 Stuttgart, DE.}

\author{Axel Arnold}
\affiliation{ICP, Institute for Computational Physics, University of Stuttgart, 
Allmandring 3, 70569 Stuttgart, DE.}

\date{\today}

\begin{abstract}
We report simulations on the homogeneous liquid-fcc nucleation of charged \mamos~for both low and high contact energy values.
As a precursor for crystal formation, we observe increased local order at the position where the crystal will form, but no correlations with the local density. Thus, the nucleation is driven by order fluctuations rather than density fluctuations.
Our results also show that the transition involves two stages in both cases, first a transition liquid~$\rightarrow$~bcc, followed by a bcc~$\rightarrow$~hcp/fcc transition.
Both transitions have to overcome free energy barriers, so that a spherical bcc-like cluster is formed first, in which the final fcc structure is nucleated mainly at the surface of the crystallite. This means that the bcc-fcc phase transition is a heterogeneous nucleation, even though we start from a homogeneous bulk liquid.

The height of the bcc~$\rightarrow$~hcp/fcc free energy barrier strongly depends on the contact energies of the \mamos.
For low contact energy this barrier is low, so that the bcc~$\rightarrow$~hcp/fcc transition happens spontaneously.
For the higher contact energy, the second barrier is too high to be crossed spontaneously by the colloidal system. However, it was possible to ratchet the system over the second barrier and to transform the bcc nuclei into the stable hcp/fcc phase.
The transitions are dominated by the first liquid-bcc transition and can be described by Classical Nucleation Theory using an effective surface tension.
\end{abstract}

\maketitle

\section{Introduction}
The crystallization of charged macromolecules plays an important role in many fields, such as biology,
soft matter physics or materials science. For example, the crystallization of proteins is required for structure determination by scattering~\cite{mcpherson04,vekilov10,kayitmazer13}. The arguably most prototypic model system consists of colloidal particles, which can be well described as charged spheres. Systems of charged colloidal particles have been used not only to model protein crystallization, but also collective properties of other systems, e.g. DNA self-assembly~\cite{stradner04}.

Experiments on colloidal systems are comparatively easy to carry out, due to their well characterized properties and the fact that they can be investigated by conventional microscopy or scattering methods~\cite{palberg99,brunner02}. Colloids, just as the smaller proteins, are charged due to dissociation, which becomes noticable under deionized conditions. Various studies under such conditions have been performed, e.g. concerning nucleation rates for different densities~\cite{schope01,schope02,wette07}.
The interaction of charged \mamos~can be well represented by a screened Coulomb or Yukawa potential~\cite{brunner02}. The phase diagram for these interactions is known~\cite{hynninen03} and exhibits two stable solid phases, namely bcc and fcc crystals.

Computer simulations using such a Yukawa interaction have been used to study the dynamics of the crystallization process and its onset, nucleation.
It has been reported that the nucleation of Yukawa particles is a two-stage process, comprising the establishment of an fcc-like core inside a previously grown bcc-like structure~\cite{auer02}. Another important question is for the precursors of nucleation. It is not yet known, whether initial clusters with a higher density are being formed first, or whether there is an increase in structural ordering which triggers the onset of crystal growth~\cite{schilling10,russo12a,tan14,granasy14}.

To investigate this, nucleation experiments and simulations at low supersaturations are necessary, where the attachment rate of growth units is slow and the nucleation process can be studied directly in great detail. The problem is that according to Classical Nucleation Theory (CNT), the free energy as a function of cluster radius $R$ is
\begin{eqnarray}
 \Delta G(R)=4\pi \gamma R^2 -\frac{4}{3}\pi \rho \Delta \mu R^3
 \label{eq:cnt}
\end{eqnarray}
with supersaturation $\Delta \mu$, surface tension $\gamma$ and number density $\rho$. This free energy exhibits a single maximum, which gives raise to a free energy barrier
\begin{equation}
\Delta G^*\equiv \Delta G(R^*)=\frac{16\pi \gamma^3}{3(\rho \Delta \mu)^2}
\end{equation}
at a critical cluster size of $R^*=2\gamma / (\rho \Delta \mu)$. A cluster smaller than this size is more likely to dissolve rather than to grow into a crystal domain. At low supersaturation $\Delta \mu$, the critical cluster and the free energy barrier are rather large, so that it takes a long waiting time until a critical cluster can be observed, which makes the process difficult to access in experiments and simulations.

CNT only assumes the existence of an initial and a final state. However, intermediate states can exist on the crystallization pathway. According to Ostwald, the phase which is closest in free energy is nucleated first, which doesn't have to be the truly stable phase~\cite{ostwald}. 
In addition, Stranski and Totomanow found that the phase with the lowest free energy barrier is nucleated first~\cite{stranski33}, and according to Alexander and McTague, the nucleation of a bcc-like phase is favored in a liquid~\cite{alexander78}.

In this work, we report simulations of nucleation in the Yukawa model for \mamos~at low supersaturations using Forward Flux Sampling (FFS) simulations. This allows us to directly investigate the homogeneous nucleation, which is triggered only by spontaneous fluctuations of the homogeneous bulk liquid. In the following section, we introduce our model and simulation details. Subsequently, we report results on nucleation pathways, precursors, and the free energy landscape compared to CNT. We conclude with a discussion of our findings.

\section{Background}

\subsection{Colloid model and simulation details}
We perform Molecular Dynamics (MD) simulations using the software package ESPResSo~\cite{espresso}. Our system consists of point particles in a 3D simulation box with periodic boundary conditions in the isothermal-isobaric ensemble, realized by a Langevin thermostat and a barostat~\cite{kolb99a}.
The particles interact via a screened Coulomb or Yukawa potential, which mimics the electrostatic interaction of weakly charged colloidal particles~\cite{auer02}. The excluded volume is modeled by a Weeks-Chandler-Andersen (WCA) potential~\cite{weeks71a}. This results in a full interaction $U(r) = U_\text{Yukawa}(r) + U_\text{WCA}(r)$ with
\begin{eqnarray}
 U_\text{Yukawa}(r) & = & \epsilon \frac{\exp(-\kappa(r/\sigma-1))}{r/\sigma},~\text{and}\\
  U_\text{WCA}(r) & = &
\begin{cases}
 4 \epsilon_\text{w}\left[ \left( \frac{\sigma}{r} \right)^{12} - \left( \frac{\sigma}{r} \right)^{6} + \frac{1}{4} \right] & r < \sigma^\frac{1}{6} \\
 0 & \text{else},
 \end{cases}
\label{eq:yukawa}
\end{eqnarray}
where $\sigma$ is the particle diameter, $\epsilon$ is the contact energy value of the Yukawa potential, $\epsilon_\text{w}$ is the contact value of the WCA potential and $\kappa$ is the inverse screening length. In this work, we set $\sigma=1$ and $\epsilon_w=1$, which determines our energy and length scales. We also fix $\kappa=5$, but 
consider two different contact energy values $\epsilon=2$ and $\epsilon=20$. The pressure was fixed at $P=25.72$ and $P=25.37$, respectively, which corresponds to two points in the stable fcc region of the phase diagram for this system~\cite{azhar00,hynninen03} with a comparable, small distance to the liquid-fcc coexistence line. We start our simulation in a metastable liquid and drive it towards crystallization using forward flux sampling (FFS) as described below. Table~\ref{tb:details} summarizes the properties of our system in the initial liquid and final solid phases.

\begin{table}
\begin{tabular}{c|c|c|c|c|c}
$\epsilon$ & $P$ & $\rho_A$ & $\rho_B$ & $\eta_A$ & $\eta_B$\\ \hline
$2$ & $25.72$ & $0.9739$ & $1.0103$ & $0.5099$ & $0.5290$ \\
$20$ & $25.37$ & $0.5618$ & $0.5668$ & $0.2942$ & $0.2968$ \\
\end{tabular}
\caption{Pressure $P$, number density $\rho$, and volume fraction $\eta$ of the liquid state $A$ and the solid state $B$, where more than $90\%$ of the particles are in a solid-like environment.}
\label{tb:details}
\end{table}

\subsection{Order parameter}
To characterize the progress of crystallization we define the size of the
largest solid cluster $\lambda=\lambda_6$ as the order parameter. We consider configurations
with $\lambda_6 < \lambda_A=5$ as liquid and systems with $\lambda_6 >
\lambda_B=7300$ as solid, where the latter value means that approximately
$90\%$ of all particles are in the largest cluster. Two particles are considered to belong to the same cluster, if they are in a solid-like environment and spatially closer than $r_\text{th}=1.47$, which is the location of the first minimum of the radial distribution function.

To detect particles in a solid-like environment, we use the per-particle local bond order parameter~\cite{steinhardt83,lechner08}, defined via
\begin{eqnarray}
\overline{q}_l(i) & = & \sqrt{\frac{4\pi}{2l+1}\sum_{m=-l}^{l}|\overline{q}_{lm}(i)|^2 },~\text{with} \\
\overline{q}_{lm}(i)& = &\frac{1}{|\widetilde{N}_b(i)|}\sum_{k\in \widetilde{N}_b(i)} q_{lm}(k)~\text{and} \\
q_{lm}(k)& = &\frac{1}{|N_b(k)|}\sum_{j\in N_b(k)}Y_{lm}(r_{kj}).
\end{eqnarray}
$q_{lm}(k)$ is a complex vector based on the spherical harmonics $Y_{lm}$ of order $l$, $N_b(k)$ is the number of nearest neighbors of particle $k$ and $r_{kj}$ is the distance between \mamos~$k$ and $j$. $\widetilde{N}_b(i)$ is the number of neighbors of the second shell.

In this work, we label a particle as solid if $\overline{q}_6(i) > q_{6,\text{th}}=0.29$. In addition, we label a particle as being in a fcc-like environment if $\overline{q}_4 > q_{4,\text{th}}=0.1$, and define a corresponding size $\lambda_4$ of the largest fcc-like cluster. Note that the measured dimensions of the crystal cluster depend sensitively on these thresholds. Our values are chosen according to the separating domains of the structures. See refs.~\cite{steinhardt83,lechner08} for further details.

\begin{figure*}[tb]
\centering
\includegraphics[width=0.99\linewidth]{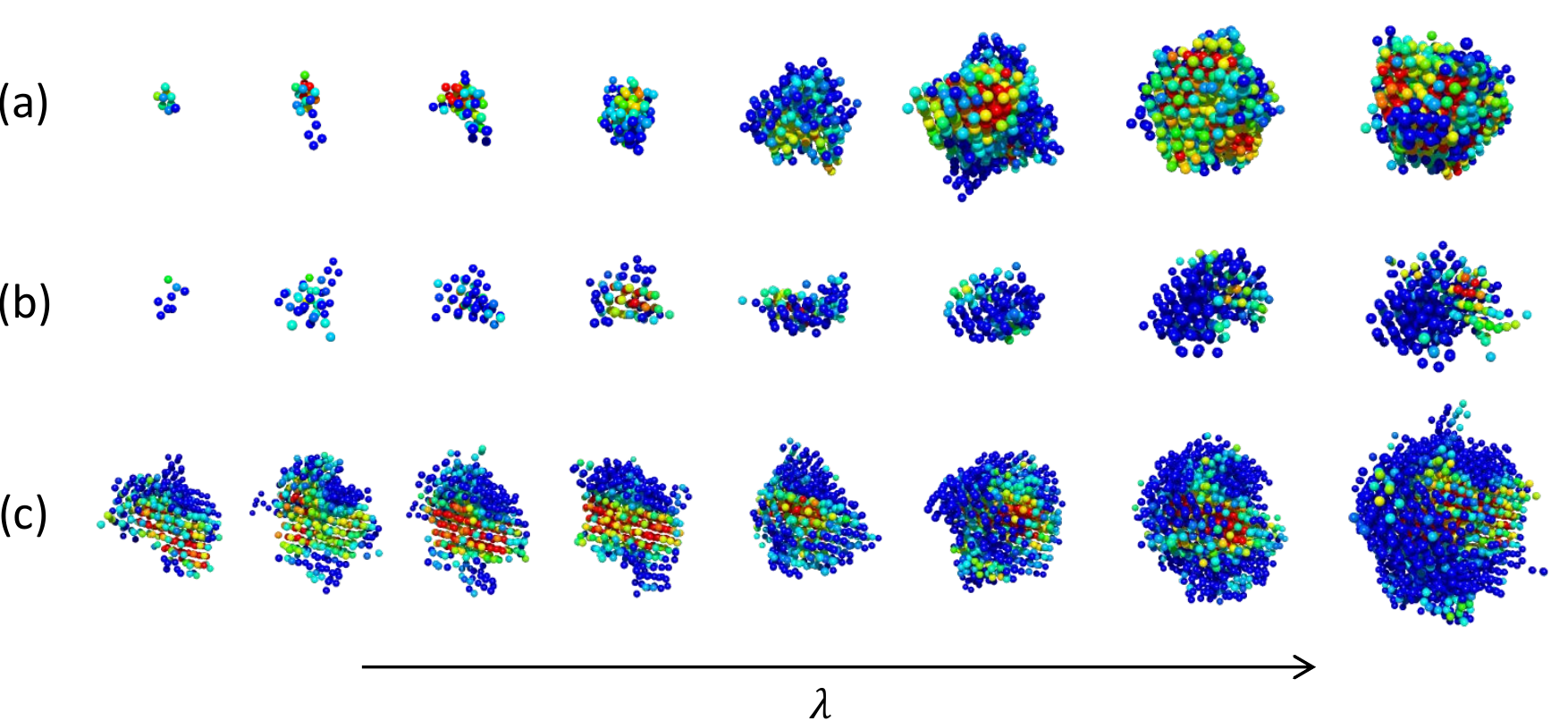}
\caption{Real-space nucleation snapshots of the \mamos: (a) Nucleation for $\epsilon=2$, (b) nucleation for $\epsilon=20$, (c) nucleation of fcc-like \mamos~starting with snapshots from the critical interface of the $\epsilon=20$ case. Blue spheres represent bcc-like \mamos~with $\overline{q}_4\leq0.05$, red spheres are fcc-like with $\overline{q}_4>0.1$, and the colors in between represent hcp-like \mamos.}
\label{fig:real_cluster}
\end{figure*}

\vspace{1cm}

\subsection{Forward Flux Sampling}
Due to the low supersaturation $\Delta \mu$ at the phase points studied, crystallization is a rare event, which cannot be studied by conventional brute-force simulations. We therefore use the Forward Flux Sampling (FFS)~\cite{allen06,allen09} technique, which is based on partitioning the transition from the initial state $A$ (liquid) to the final state $B$ (solid) into multiple stages. The transition is characterized in terms of the order parameter $\lambda$, with $\lambda \leq \lambda_0\equiv \lambda_A$ if the system is in $A$, $\lambda \geq \lambda_n\equiv \lambda_B$ if in $B$, and $\lambda_A <\lambda <\lambda_B$ for the transition region.   The partitioning is performed via a set of $n$ interfaces at discrete positions $\lambda_i$.
This means that the overall transition is split up into the calculation of smaller trajectory fragments for each stage, which are much more likely to occur.

The transition rate from state $A$ to state $B$ is then calculated by
\begin{equation}
 k_{AB}=\Phi P_B,
\end{equation}
where $\Phi$ is the so-called escape flux for leaving $A$, and $P_B=\prod_i{p_i}$ the overall probability to reach $B$ coming from $\lambda_A$ by crossing the interfaces with their respective transitions probabilities $p_i$.

We use the direct FFS (DFFS) algorithm to calculate $\Phi$ and $P_B$. $\Phi=N_0/t$ is calculated by starting an initial MD simulation run from a random phase point in $A$ and counting the number $N_0$ of $\lambda_A$ crossings in positive $\lambda$-direction during a simulation time $t$. In addition, a system configuration snapshot is stored on each crossing.

In the next step, previously stored configurations on the last interface $\lambda_i$ are drawn at random and trial runs are launched starting from these points, with the possibility to either fall back to $\lambda_A$ (`failure') or to reach the next interface $\lambda_{i+1}$ (`success'). This results in the transition probabilities $p_i=M_i/M$, with the number of overall trials $M$ and the number of successful trials $M_i$.

In our simulations, we only specify the border of the states $\lambda_A$ and $\lambda_B$. The interface set $\lambda_i$ is determined automatically and optimized during simulation using the exploring scout placement method from~\cite{kratzer13}, with target transition probability $p_\text{des}=0.5$ and $20$ exploratory runs per interface. For the implementation and parallelization we use the Flexible Rare Event Sampling Harness System (FRESHS)~\cite{kratzer14}.

With FFS, not only transition rates can be calculated, but also the physical pathways can be investigated by backtracking the successful runs.
In addition, the free energy profile
\begin{equation}
 \Delta G(\lambda) \propto -k_BT\log{[\rho(\lambda)]}
\end{equation}
along the reaction coordinate $\lambda$ can be obtained by splitting up the calculation of the stationary distribution $\rho(\lambda)$ into a forward and a backward contribution $\Psi$~\cite{valeriani07}, 
\begin{equation}
 \rho(\lambda)=\Psi_A(\lambda)+\Psi_B(\lambda),
 \label{eq:rho_psi_AB}
\end{equation}
which in our case means that for sampling the backward contribution $\Psi_B(\lambda)$, we dissolve the crystallite cluster again in a reverse simulation.

\vspace{1cm}

\section{Results}

\subsection{Nucleation Pathways}

Fig.~\ref{fig:real_cluster}(a+b) show exemplary nucleation paths obtained from FFS, driving along the largest solid cluster size $\lambda_6$.  A summary of the simulation details is given in Appendix~\ref{sec:appendix_FFS}. As it can be seen, the nucleating clusters are almost spherical for both employed contact energies of the interaction potential. This is expected from CNT at these low supersaturations implying comparatively large surface tensions.

For a low contact energy of the colloids, $\epsilon=2$, we indeed observe a transition to a solid phase with a strong fcc-like signature as expected from the phase diagram. This cannot be seen immediately from Fig.~\ref{fig:real_cluster}(a), because the fcc-like kernel is covered by bcc or hcp-like surface layer.

\begin{figure}[tb]
\centering
\makebox[20pt][l]{(a)}{\rotatebox{0}{{\includegraphics[width=0.68\linewidth]{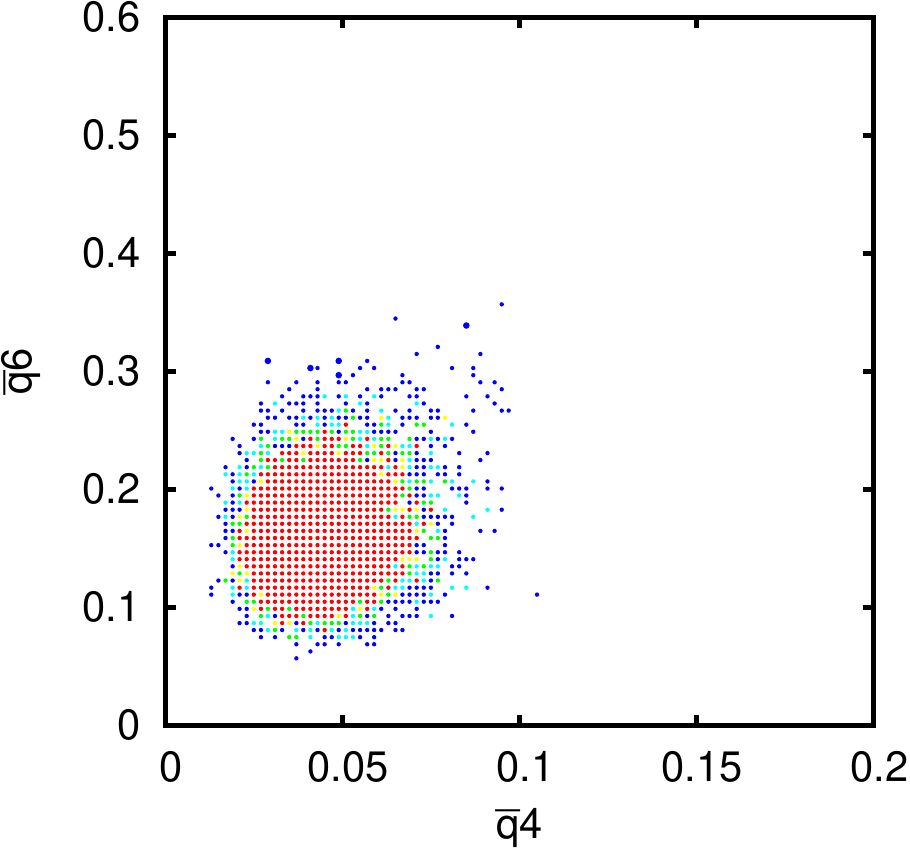}}}}\vspace{0.5cm}
\makebox[20pt][l]{(b)}{\rotatebox{0}{{\includegraphics[width=0.68\linewidth]{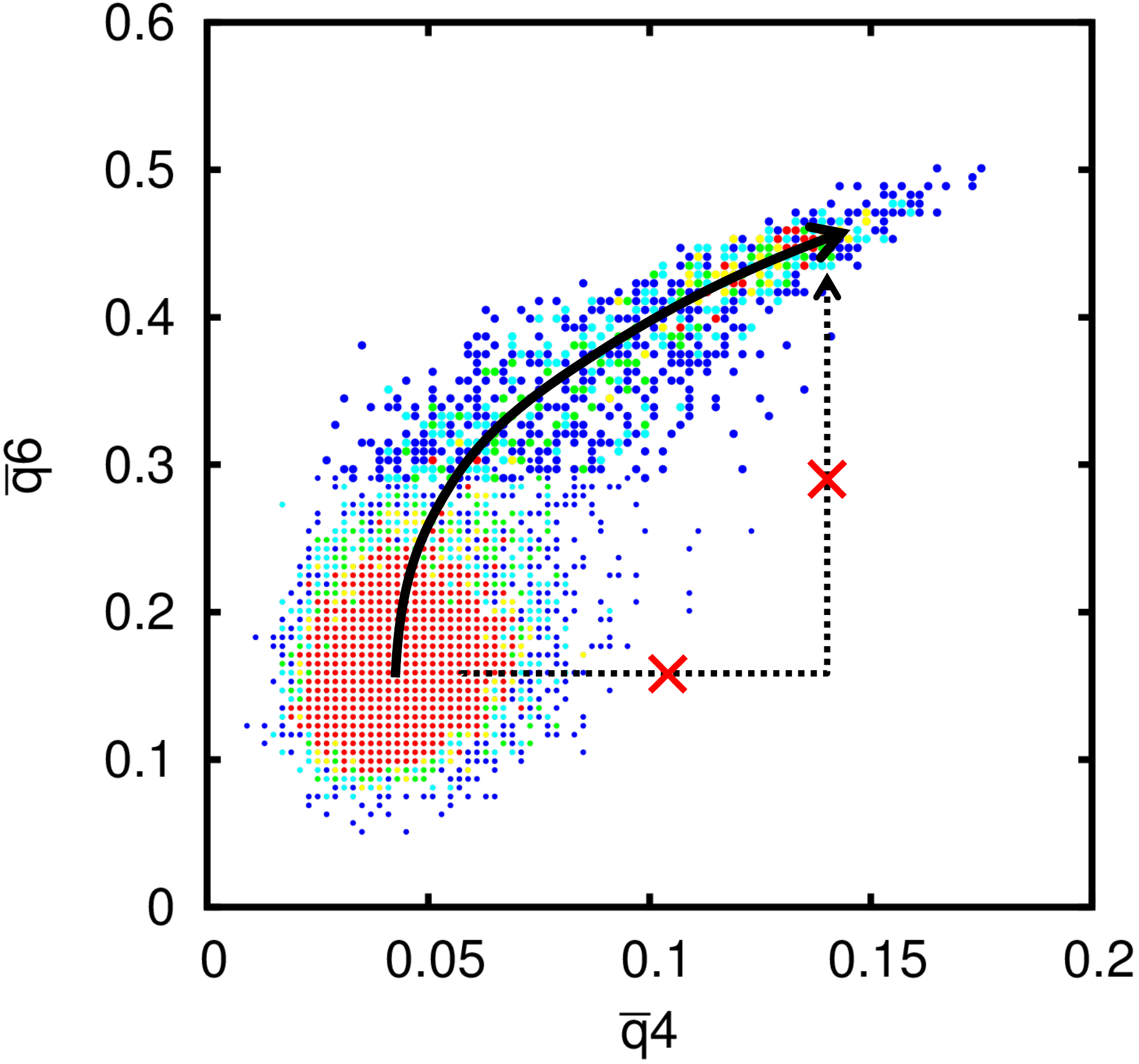}}}}\vspace{0.5cm}
\makebox[20pt][l]{(c)}{\rotatebox{0}{{\includegraphics[width=0.68\linewidth]{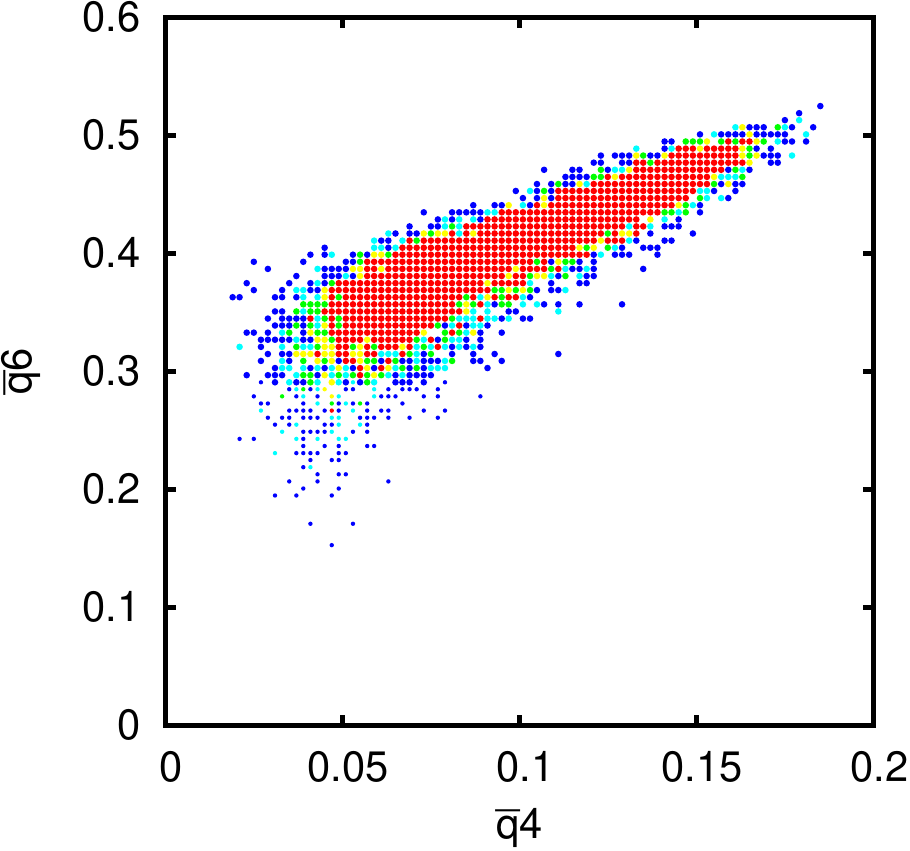}}}}
\caption{$\overline{q}_4\overline{q}_6$-panel of the transition pathway for $\epsilon=2$. The larger dots represent particles which are member of the largest cluster. Blue color coding stands for $1$ particle and red color coding for $5$ particles having a similar $(\overline{q}_4,\overline{q}_6)$ value pair. Shown are the snapshots on (a) $\lambda_0$ (border of $A$), (b) $\lambda_{n-1}$ (critical cluster) and (c) $\lambda_n$ (border $\lambda_B$).}
\label{fig:panel2}
\end{figure}

Fig.~\ref{fig:panel2} shows a selection of $\overline{q}_4\overline{q}_6$-panels of the transition for $\epsilon=2$. At $\lambda_0$, the system is mainly in the liquid state, with only a few particles being solid-like. During nucleation of solid particles ($\overline{q}_6\geq0.29$) also the $\overline{q}_4$ order parameter increases, indicating a strong tendency to a hcp/fcc-like structure. In the final state, where the main fraction of particles reside in the solid cluster, also most particles are characterized as fcc-like.
The observed pathway is therefore:
\begin{equation*}
 \text{liquid} (\rightarrow \text{bcc}) \rightarrow \text{hcp/fcc}.
\end{equation*}
We do not observe a direct transition to the hcp/fcc-like state without crossing the bcc-like domain at lower $\overline{q}_4$ values, which is indicated by the crossed-out arrow in Fig.~\ref{fig:panel2}(b). The overall transition proceeds spontaneously with a mixture of bcc-like, hcp-like and fcc-like particles. This can also be seen in the real-space nucleation trace in Fig.~\ref{fig:real_cluster}(a), where the last snapshot corresponds to Fig.~\ref{fig:panel2}(b). There is no retardation or stay in an intermediate stage, only a crossing of the bcc-like domain (as indicated by the parentheses in the above scheme sequence). 

At high contact energy $\epsilon=20$, none of the pathways generated by driving along $\lambda_6$ reached the thermodynamically stable fcc structure. The solid clusters rather grow spontaneously into bcc crystals, as shown exemplarily in Fig.~\ref{fig:real_cluster}(b).

In order to investigate the transition to the thermodynamically stable fcc phase, we used our FFS simulations to explicitly drive the system towards a larger cluster of fcc-like particles. To this aim, we used a different order parameter, namely the largest cluster of particles with fcc-like environment, as detected by $\overline{q}_4$. If started again from the bulk liquid, FFS is not able to drive cluster growth, indicating that $\lambda_4$ is not a good reaction coordinate for this nucleation. While FFS does not require the order parameter to be a reaction coordinate~\cite{allen09}, it becomes very ineffective if the order parameter is not correlated with the actual reaction coordinate. What we therefore observe is that fourfold symmetry alone does not allow for stable solid clusters, which we will confirm later. The transition of a bulk bcc phase to a bulk fcc phase also involves a large energy barrier and thus is also extremely unlikely to happen. Hence, we can conclude that there exist no transitions of the form
\begin{eqnarray*}
 \text{liquid} & \nrightarrow & \text{hcp/fcc}, \text{ or}\\
 \text{bcc} & \nrightarrow & \text{hcp/fcc}.
\end{eqnarray*}

\begin{figure*}[tb]
\centering
\begin{tikzpicture}[auto]
\node [eblock] (liquid) {\pgfbox[center,bottom]{\pgftext{\includegraphics[width=\tps]{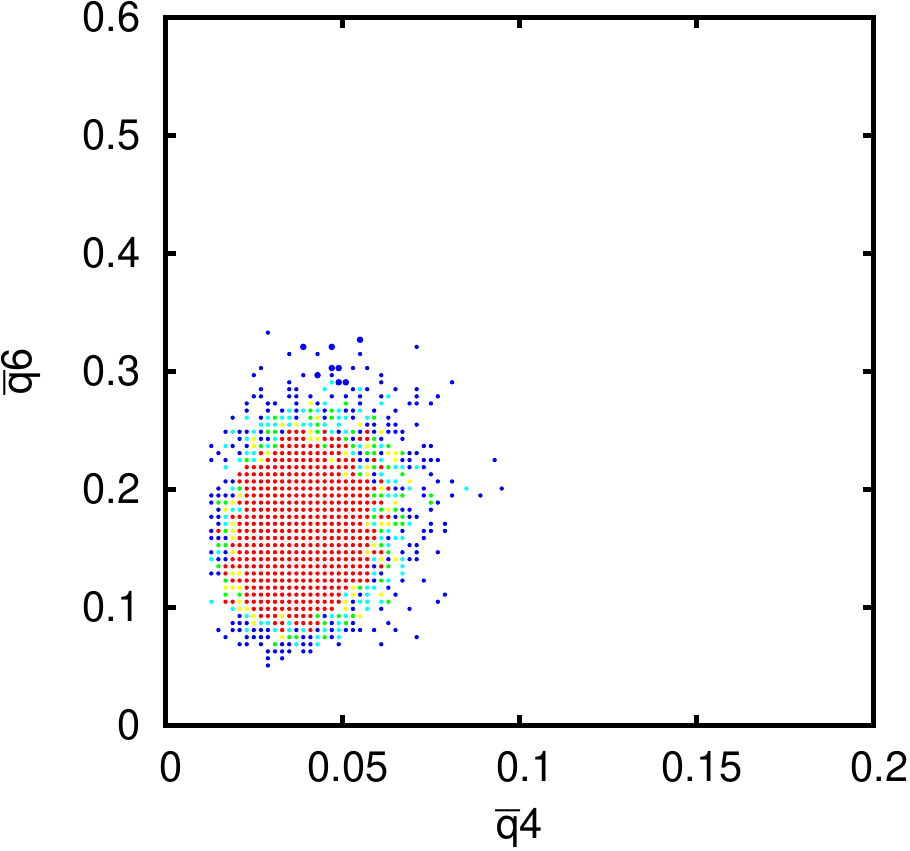}}}};
\node [eblock,right of=liquid,node distance=\tndv] (critical) {\pgfbox[center,bottom]{\pgftext{\includegraphics[width=\tps]{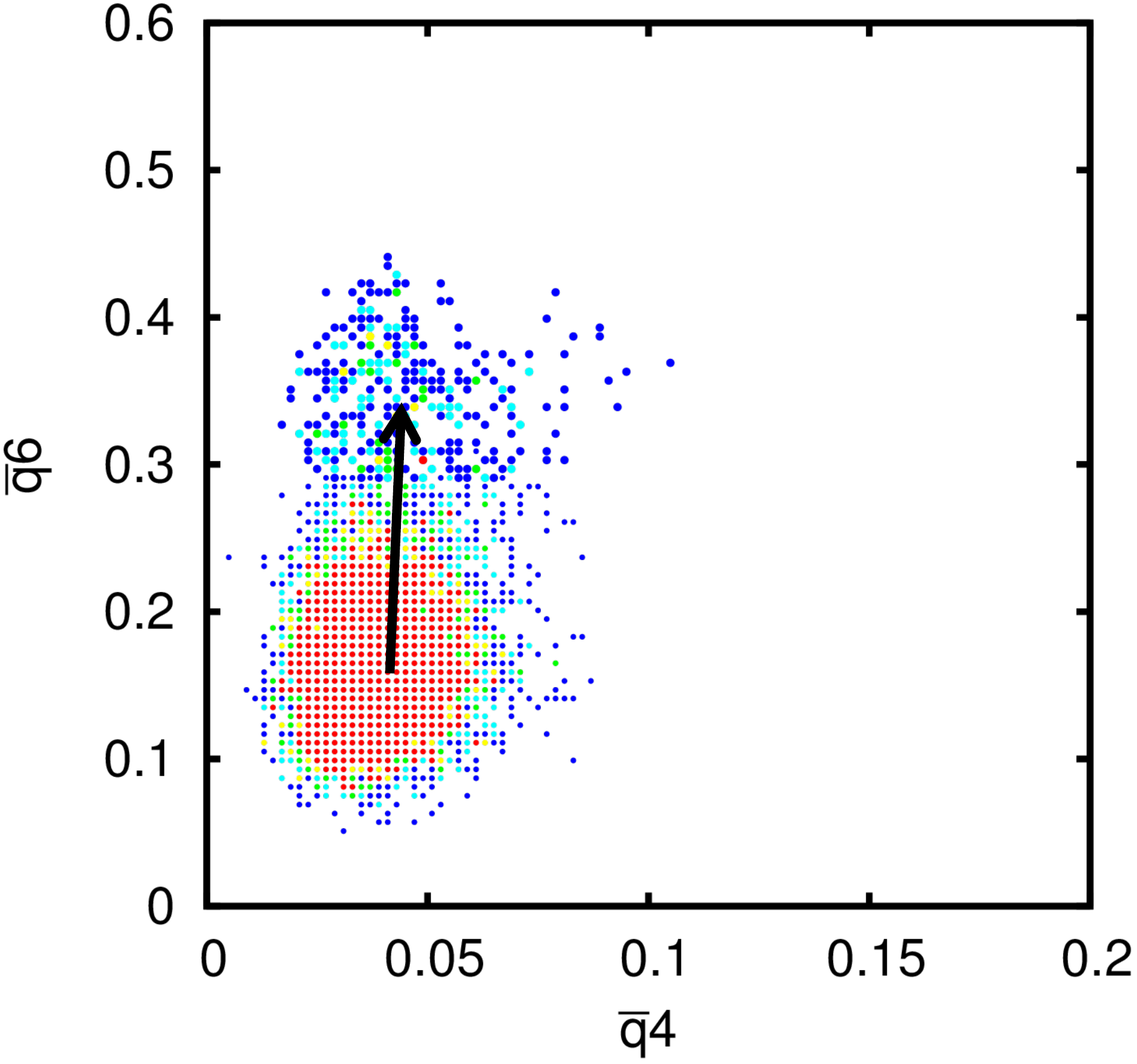}}}};
\node [eblock,right of=critical,node distance=\tndv] (final) {\pgfbox[center,bottom]{\pgftext{\includegraphics[width=\tps]{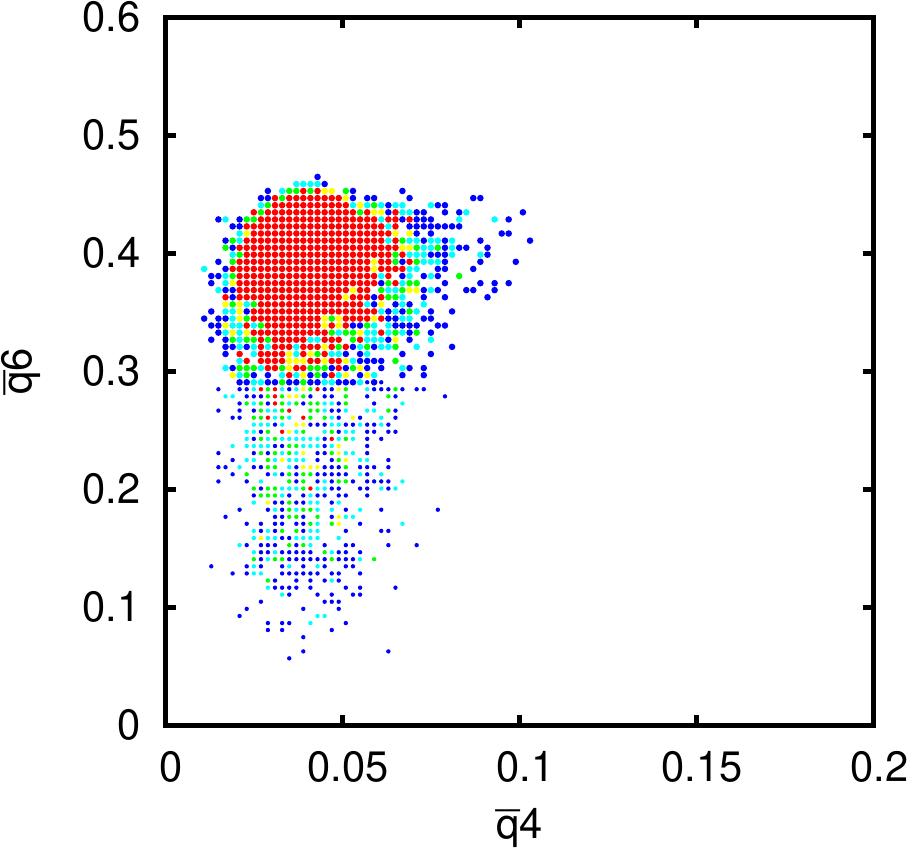}}}};
\node [eblock,below of=liquid,node distance=\tndv] (fcc1) {\pgfbox[center,bottom]{\pgftext{\includegraphics[width=\tps]{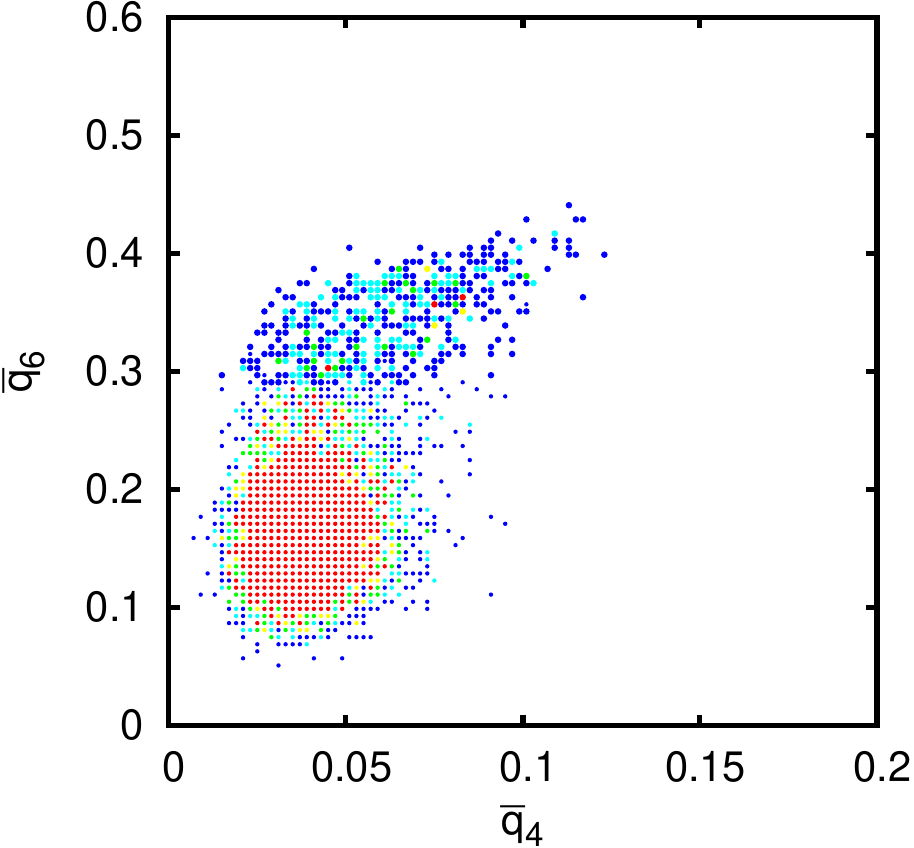}}}};
\node [eblock,right of=fcc1,node distance=\tndv] (fcc2) {\pgfbox[center,bottom]{\pgftext{\includegraphics[width=\tps]{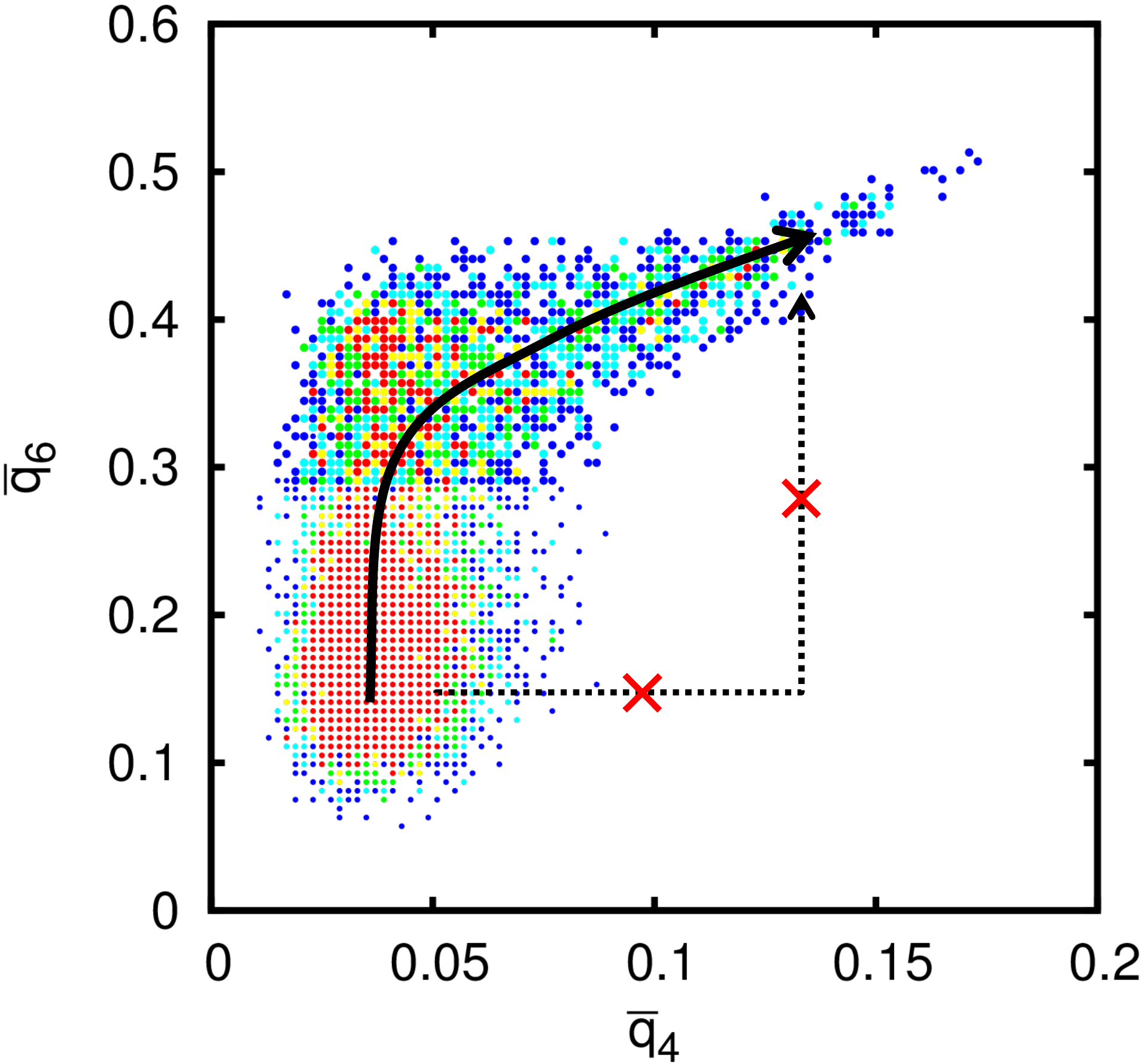}}}};
\node [eblock,right of=fcc2,node distance=\tndv] (fcc3) {\pgfbox[center,bottom]{\pgftext{\includegraphics[width=\tps]{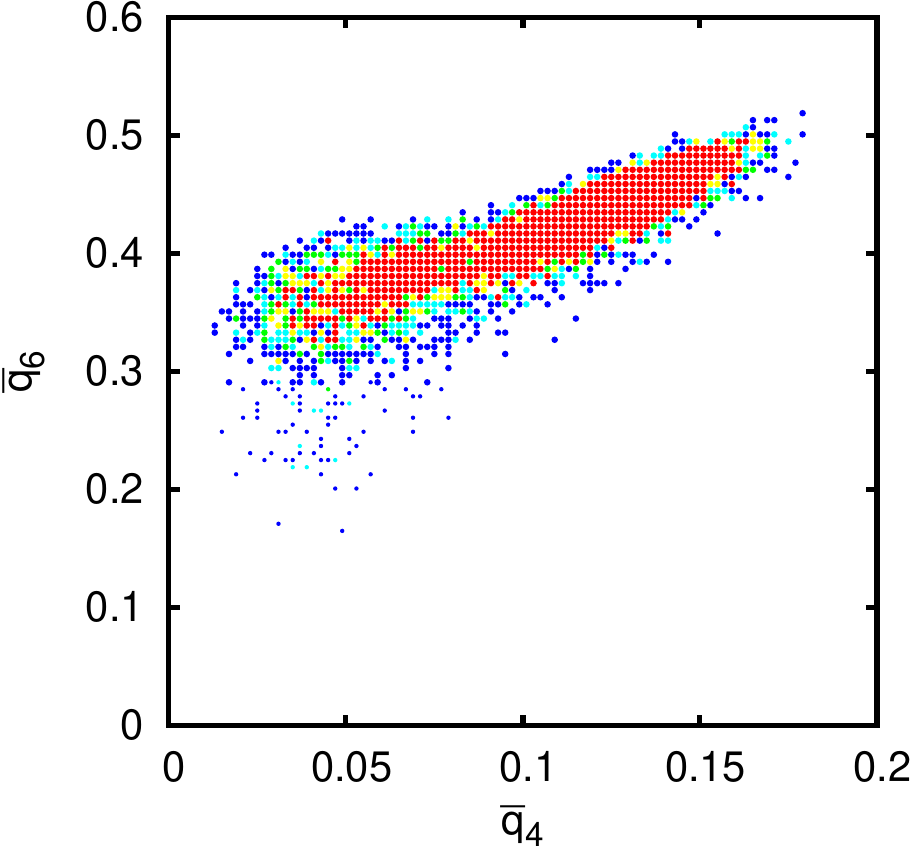}}}};
\coordinate [below of=liquid, node distance=\tndh] (hidden) {};
\path [line] (liquid) -- node {\qsix} (critical);
\path [line] (critical) -- node {\qsix} (final);
\path [line] (critical) |- (hidden) node [near end,above] {\qfour} -| (fcc1);
\path [line] (fcc1) -- node {\qfour} (fcc2);
\path [line] (fcc2) -- node {\qfour} (fcc3);
\end{tikzpicture}
\caption{$\overline{q}_4\overline{q}_6$-panels of the transition pathways for $\epsilon=20$. The first row shows the spontaneous transition to the bcc-like state, with snapshots on $\lambda_0$ (border of $A$), $\lambda_{n-1}$ (critical cluster) and $\lambda_n$ (border $\lambda_B$). The second row begins at the system state comprising the critical cluster and undergoes a transition to the hcp/fcc-like state.}
\label{fig:panel20}
\end{figure*}
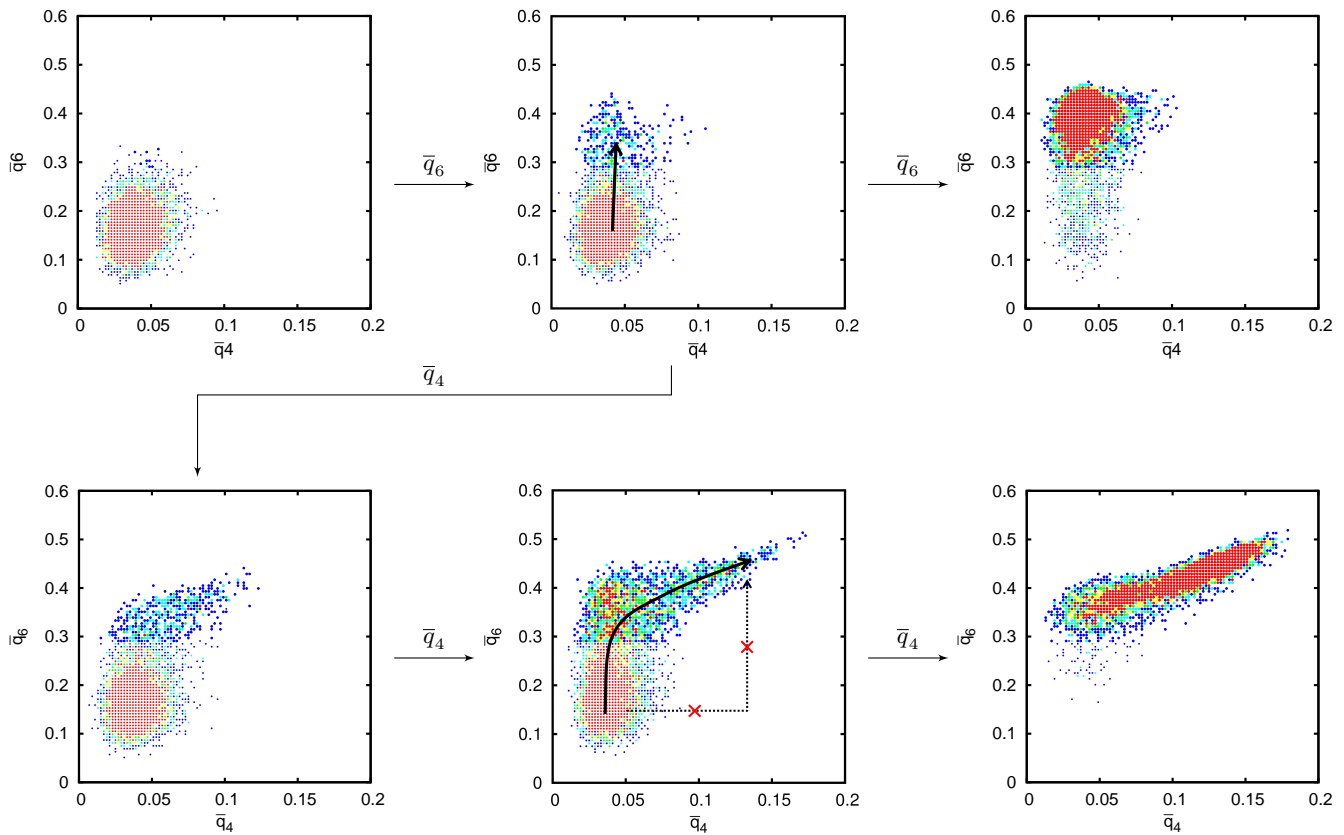

However, it is possible to grow the stable fcc phase from configurations containing critical bcc-like clusters. Such a small bcc-like cluster can transform into an fcc-cluster, which then continues to grow. We obtain these starting configurations from the second to last automatically placed FFS interface, because we chose the target transition probability to $p_\text{des}=0.5$. This means that no additional interface is placed beyond a committor value of $0.5$~\cite{bolhuis02}, which corresponds to the committor value at the critical cluster size. For example, the trace in Fig.~\ref{fig:real_cluster}(c) was obtained by starting from the rightmost configuration shown in Fig.~\ref{fig:real_cluster}(b). Effectively, this corresponds to an order parameter that is the largest solid cluster in the beginning, and, after a certain size, turns over into an order parameter that corresponds to the largest cluster of fourfold symmetry. Since for this to work the solid cluster needs to be stable, and we can start at earliest from a critical bcc cluster.

Fig.~\ref{fig:panel20} summarizes the transitions for $\epsilon=20$.
The first row shows the most likely spontaneous transition when the system crystallizes into a bcc crystal. This corresponds to the real-space trace in Fig.~\ref{fig:real_cluster}(b). Only a few \mamos~have higher $\overline{q}_4$ values, which is much less than for the $\epsilon=2$ case.  Hence, the spontaneous transition reads
\begin{equation*}
 \text{liquid} \rightarrow \text{bcc}.
\end{equation*}
The second row in Fig.~\ref{fig:panel20} shows the case where we drove the system to a higher number of fcc-like particles starting from critical bcc-clusters, that is, the transition
\begin{eqnarray*}
 \text{bcc-critical} & \rightarrow & \text{hcp/fcc}.
\end{eqnarray*}
Note that in this case the system initially gains sixfold symmetry faster than fourfold symmetry, despite our driving along $\lambda_4$. This shows that our order parameter is still not a good reaction coordinate, however in this case good enough so that FFS can enforce the fcc transition. What we observe is that the bcc cluster grows slightly above the critical size before transforming into fcc, so that the overall transition is much less smooth than for $\epsilon=2$. The overall transition can thus be characterized as
\begin{equation*}
 \text{liquid} \rightarrow \text{bcc-critical} \rightarrow \text{hcp/fcc}.
\end{equation*}

\vspace{0.8cm}

\subsection{Two-stage nucleation}

This behavior can be explained naturally if nucleation in Yukawa systems is a two-stage process with two free energy barriers, one for the first transition to the bcc-like structure and another one for the second transition to the hcp/fcc-like structure.

The height of the bcc-fcc free energy barrier is dependent on the contact value $\epsilon$. In the first case of $\epsilon=2$, the energy barrier for the transition bcc$\rightarrow$hcp/fcc can be overcome spontaneously and is therefore hidden during the nucleation process.
In contrast, for $\epsilon=20$, the system is more likely to nucleate a bcc-like critical cluster, and  performs a second transition to the stable fcc phase only at a later stage. This indicates a higher barrier than in the previous case, which in simulations can only be overcome by using rare event sampling again.

\begin{figure}[tb]
\centering
\includegraphics[width=0.9\linewidth]{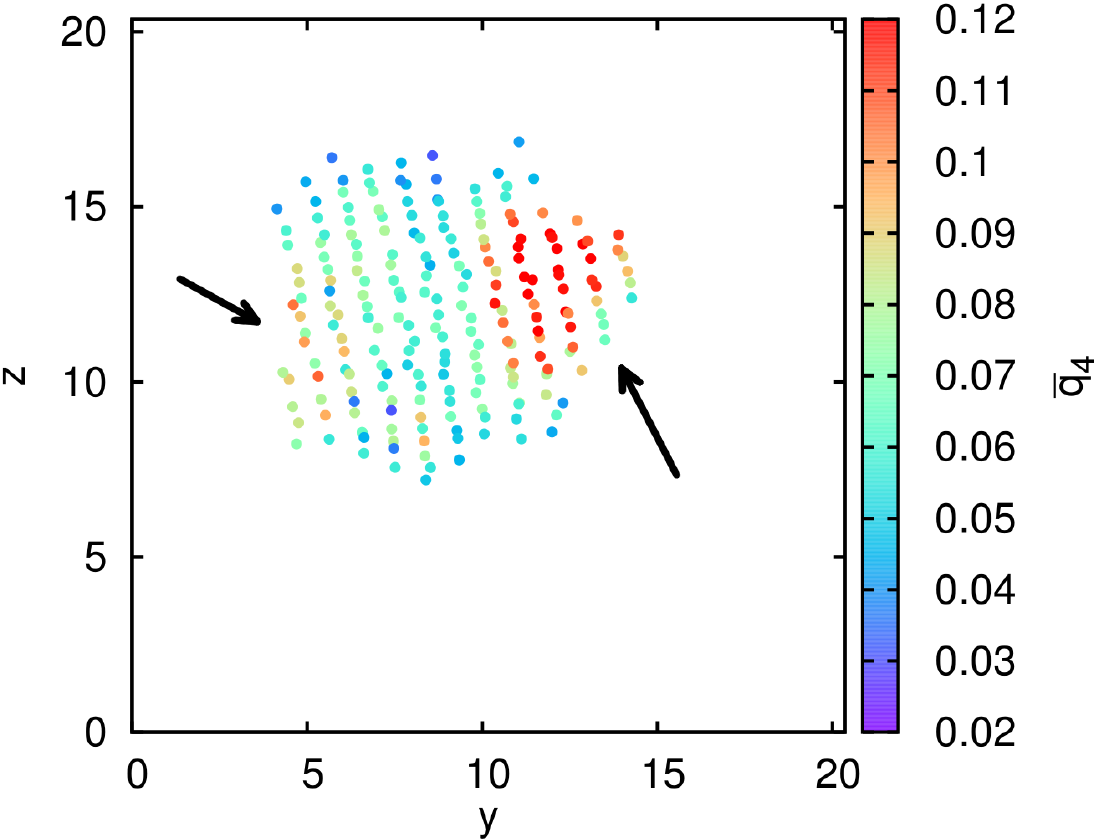}
\caption{Snapshot of a slice in $(y,z)$-direction through the crystal cluster during nucleation. The color coding is according to the $\overline{q}_4$ value. In this snapshot, the values at the border of the crystal cluster are higher than in the middle, but fluctuate during time evolution.}
\label{fig:growth_mechanism}
\end{figure}

The fact that we observe the latter transition only starting from critical bcc clusters can be understood from the $\epsilon=2$ scenario. Fig.~\ref{fig:growth_mechanism} shows a representative snapshot of a slice through the cluster during nucleation.
We observe that the main $\overline{q}_4$ fluctuations occur at the surface. Thus, the fcc structure is preferentially grown from these fluctuations at the liquid-bcc interface of the growing bcc cluster. This can be explained by assuming a free energy barrier between bcc and fcc. The corresponding nucleation event is facilitated by the presence of the liquid-bcc surface. In other words, the fcc phase is heterogeneously nucleated from the bcc phase, even though we start from a homogeneous bulk liquid. 

To quantify this process, we calculated the transition rate with respect to the surface area of the critical-bcc cluster, where the $\overline{q}_4$ fluctuations take place.
In this case, we obtain a transition rate of
\begin{eqnarray*}
k_{\text{critical-bcc},\text{fcc}} & = & 9.52\times 10^{-08}\tau^{-1}\sigma^{-2}.
\end{eqnarray*}
Further details of this transition can be found in the Appendix~\ref{sec:appendix_FFS}, Table~\ref{tb:rates_fcc}.

\vspace{1cm}

\subsection{Precursors}
The investigation of precursors is possible by backtracking the successful transition pathways to the border of state $A$, namely to the snapshots on $\lambda_0$ from which the critical clusters are nucleated.

Note that the snapshot on $\lambda_0$ is the result of the initial MD simulation run for the escape flux calculation and was sampled without driving the system with an order parameter. We ensured that the crystal cluster on $\lambda_0$ is the one from which the critical cluster is nucleated by monitoring the center of mass of the crystal cluster during the nucleation process. There are no discontinuities in the center of mass offsets relative to the box in our traces, which means that we grew only one largest cluster in the system and that there are no other clusters with a competing size.

\begin{figure*}[tb]
\centering
{\rotatebox{0}{{\includegraphics[width=0.323\linewidth,clip=true]{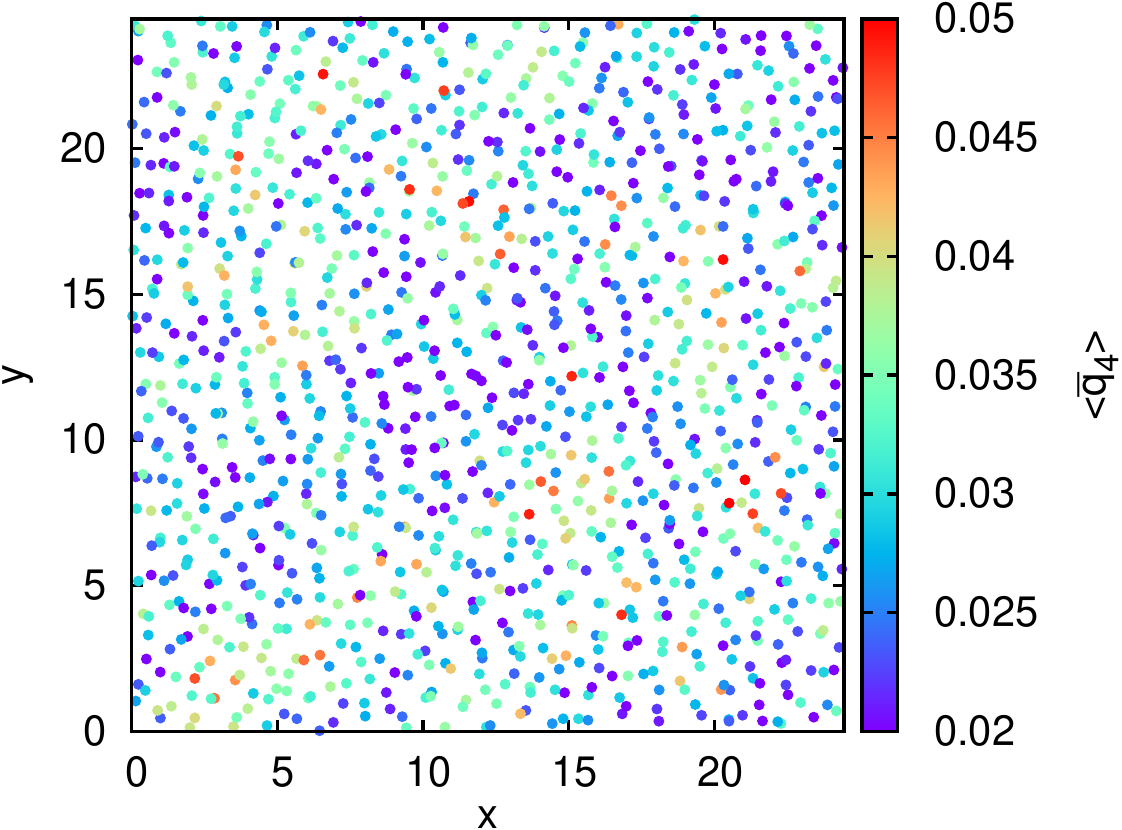}}}}
{\rotatebox{0}{{\includegraphics[width=0.321\linewidth,clip=true]{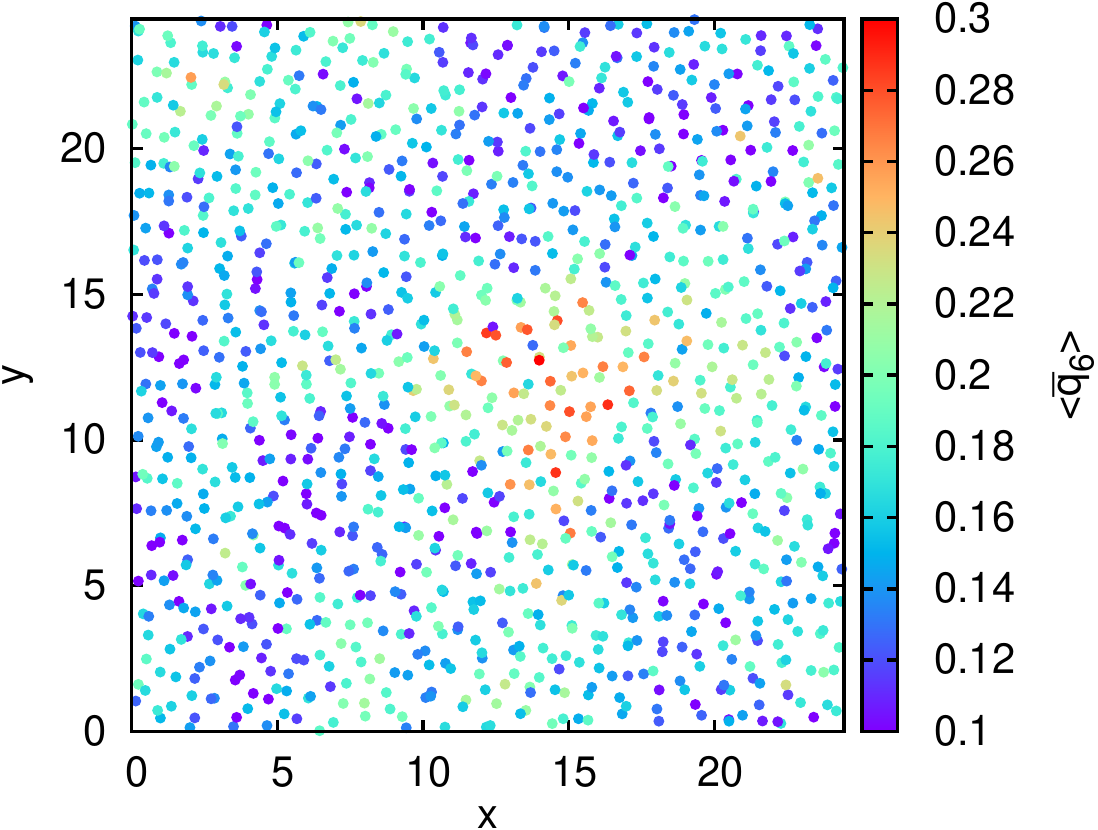}}}}
{\rotatebox{0}{{\includegraphics[width=0.32\linewidth,clip=true]{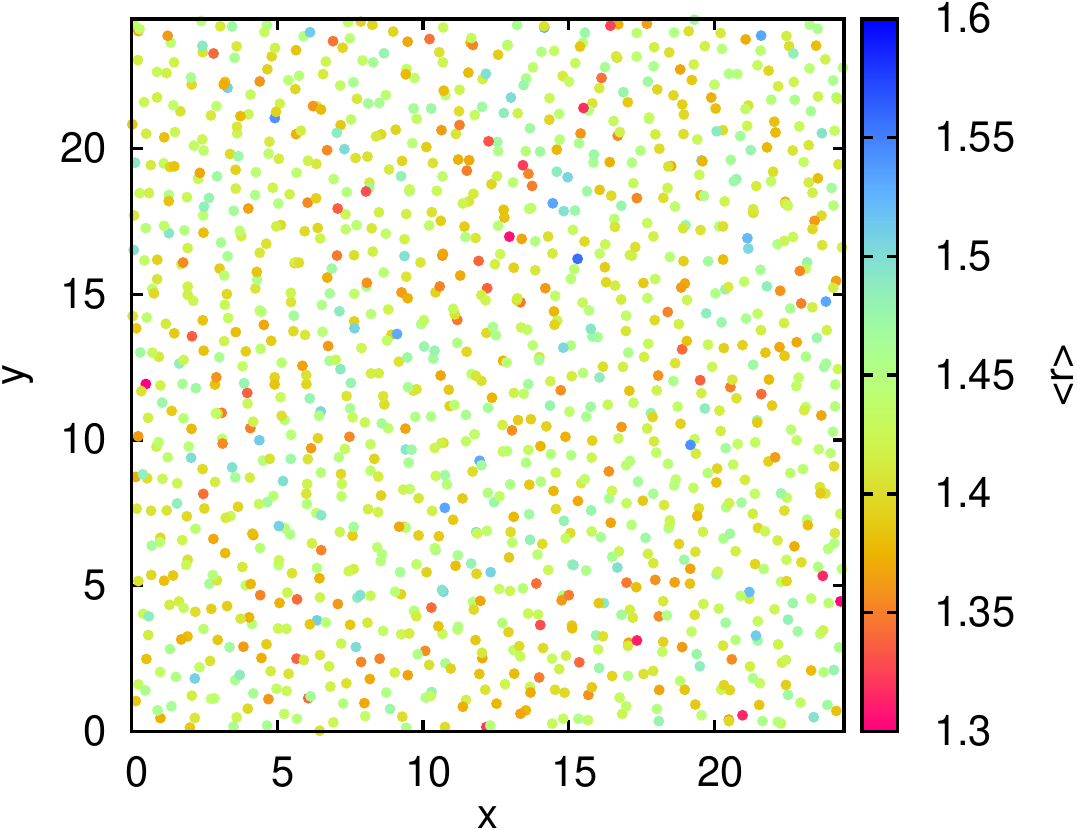}}}}
{\rotatebox{0}{{\includegraphics[width=0.323\linewidth,clip=true]{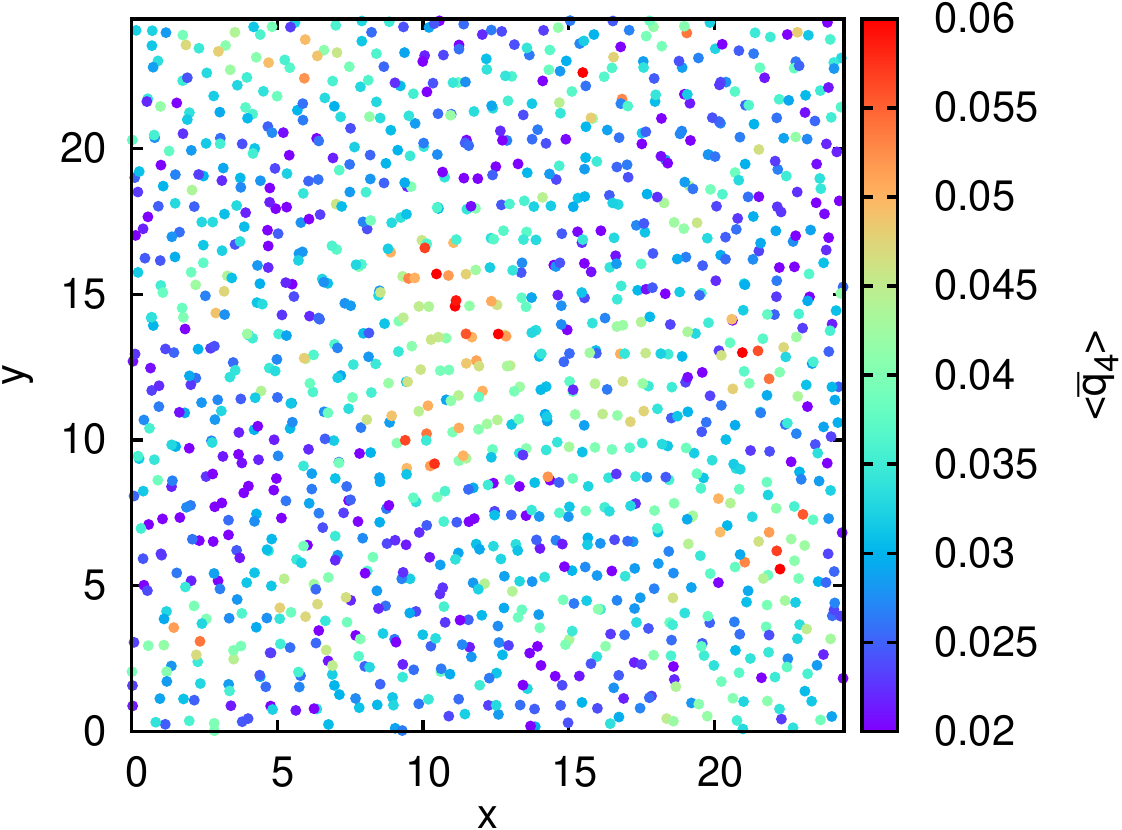}}}}
{\rotatebox{0}{{\includegraphics[width=0.321\linewidth,clip=true]{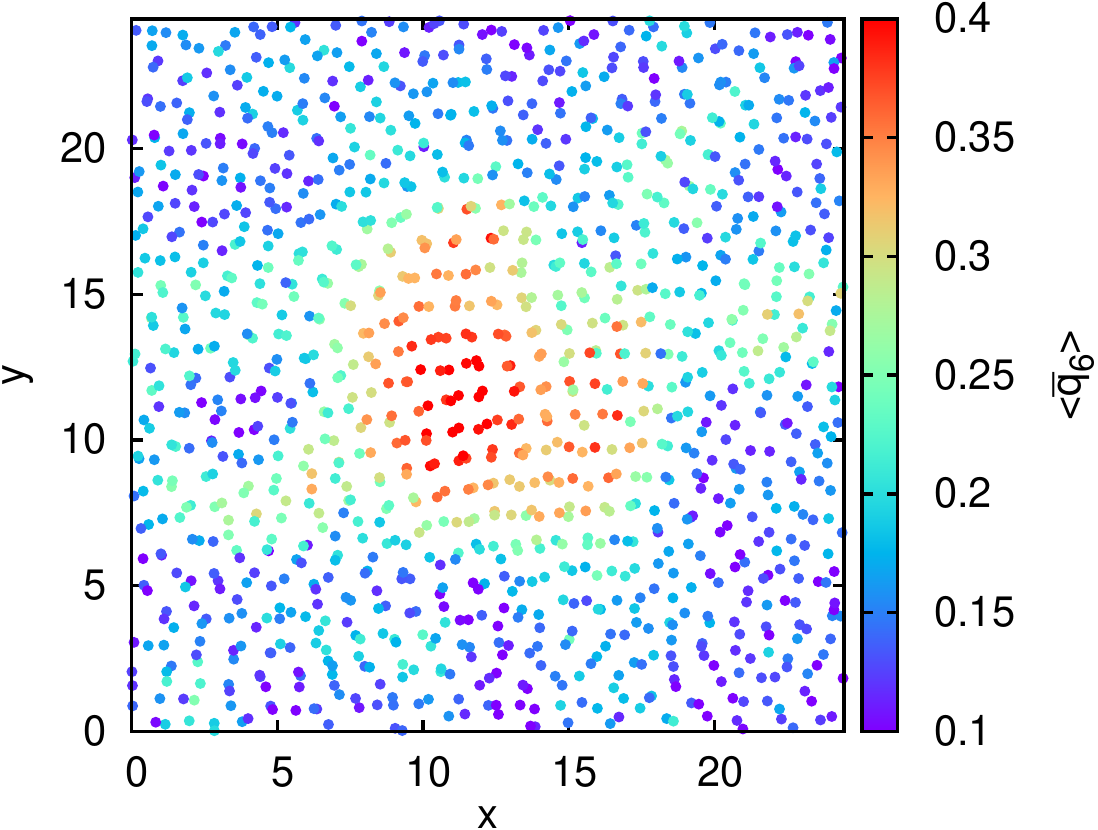}}}}
{\rotatebox{0}{{\includegraphics[width=0.32\linewidth,clip=true]{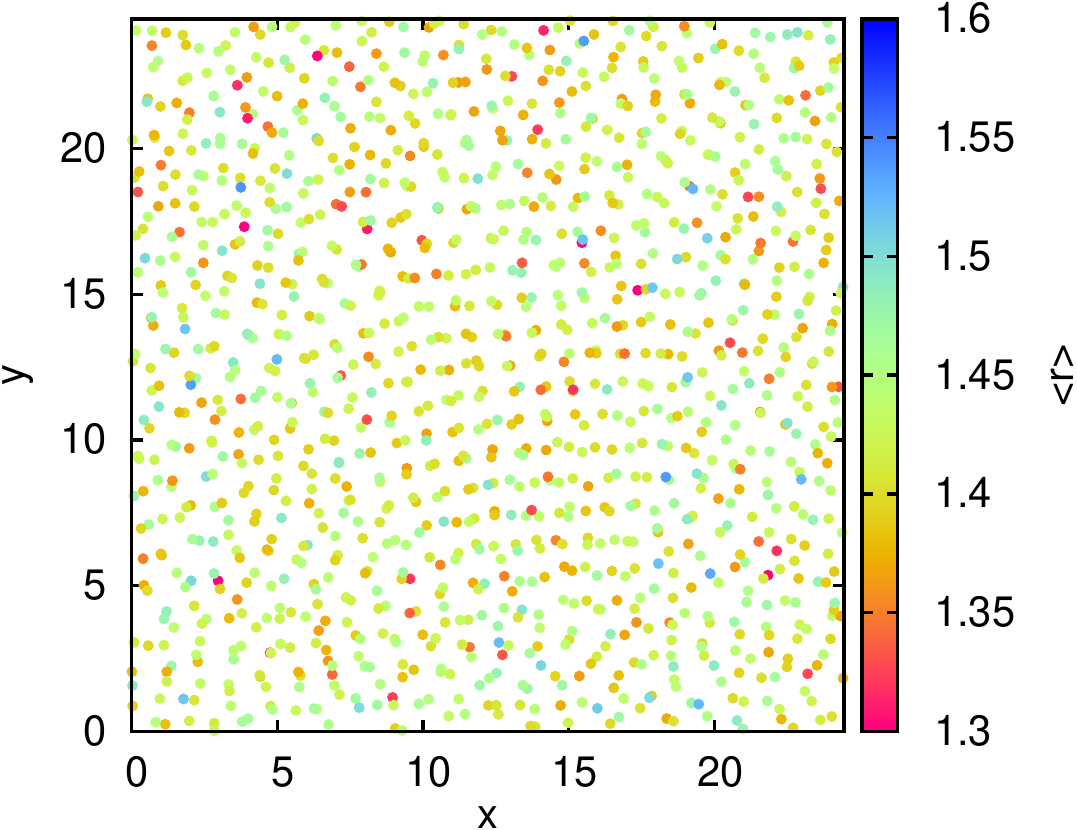}}}}
\caption{Visualizing precursors for nucleation by backtracking the successful pathways to the early stage $\lambda_0$ (first row) and comparison to the same location at the later stage of the critical cluster (second row). There are no correlations at an early stage to the position where the critical cluster forms for the $\overline{q}_4$ panel (left) and for the neighbor distance panel (right). However, the $\overline{q}_6$ panel (middle) shows already elevated values at the corresponding position.}
\label{fig:precursor}
\end{figure*}

Fig.~\ref{fig:precursor} shows slices through the 3D system at the early stage (first row) and at the stage of the critical cluster (second row). 
The panels show the $\overline{q}_4$ and $\overline{q}_6$ values for all particles as well as the averaged neighbor distances of the particles, which corresponds to a density map. For better visibility, we transformed the overall system snapshots according to the center of mass of the critical cluster for both stages.

We find that there are no correlations with the location of the critical cluster at the early stage for the $\overline{q}_4$ parameter and with the density, the values are distributed randomly over the whole domain. This a posteriori explains why nucleation cannot be driven by the $\lambda_4$ order parameter from the bulk liquid.

In contrast, the $\overline{q}_6$ order parameter shows already elevated values at the corresponding position where the critical cluster will nucleate.
Thus, the $\overline{q}_6$ distribution is the quantity of the system which shows a distinct indication at this early stage where the critical cluster will form and can be seen as a precursor in this case. As the $\lambda_A$ border is from the initial brute-force simulation run, this can not be an artifact from the FFS simulation being driven by the $\overline{q}_6$, and therefore demonstrates retrospectively that this was a good choice for an order parameter.

\vspace{1cm}

\subsection{Comparison to CNT}
For comparing to classical nucleation theory, we calculated the free energy difference $\Delta G(\lambda)$ by computing the stationary distributions $\rho(\lambda)$ according to Eq.~(\ref{eq:rho_psi_AB}) in a forward and a backward FFS simulation run. In addition, the brute-force sampling of the order parameter distribution in state $A$ was fitted by a least-square fit to the profile  as described in ref.~\cite{valeriani07}.
Note, that for $\epsilon=20$ we only report the free energy landscape for the liquid~$\rightarrow$~bcc transition. As shown, the second transition to the hcp/fcc-like state occurs much later, and that will not change the energy landscape before the critical cluster size.

\begin{figure}[tb]
\centering
\includegraphics[width=0.99\linewidth,clip=true]{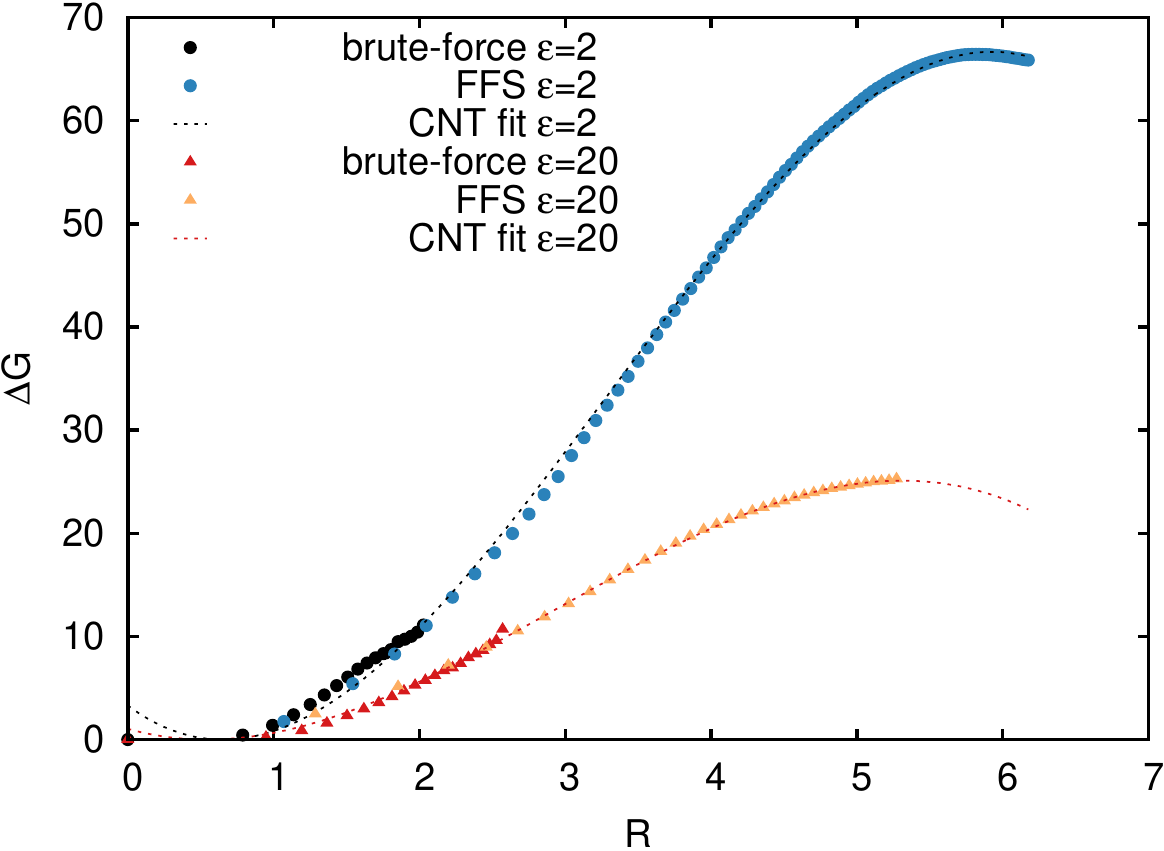}
\caption{$\Delta G$ for $\epsilon=2$ and the first transition to the critical-bcc cluster for $\epsilon=20$, obtained from FFS simulations with $\overline{q}_6\geq 0.29$ for the identification of solid particles. We allow a shift of the theory curve in $R$ direction because of the unknown real cluster size, which is dependent on the $\overline{q}_6$ threshold.
  \label{fig:dG_2_20}}
\end{figure}

Fig.~\ref{fig:dG_2_20} shows the $\Delta G(R)$ profile for $\epsilon=2$ and $\epsilon=20$, which we obtain using an effective radius $R=((3n)/(4\rho \pi))^{1/3}$ and assuming spherical cluster growth. 
Note that the simulations have been performed with a $\overline{q}_6\geq
0.29$ threshold for the identification of solid particles. This threshold
influences the cluster size because the $\overline{q}_6$ profile decays
smoothly at the border of the cluster. Therefore, we introduce a shift $\Delta R$ in the
cluster size $R$, which in our case is $\Delta R=0.65$ and $\Delta R=0.54$ for the contact energies $\epsilon=2$ and $\epsilon=20$, respectively, i.~e. less than a particle diameter. This shows that the common choice of $\overline{q}_6\geq 0.29$ is indeed suitable to describe solid nucleation.

Table~\ref{tb:fitparams} summarizes the fitting results.
\begin{table}
\centering
\begin{tabular}{c|ccccccccc}
$\epsilon$ & $R^*$ & $\Delta G^*$ & $\Delta R$ & $\Delta \mu$ & $\gamma$ & $\lambda^*$ & $\lambda_{n-2}$ & $\lambda_{n-1}$ \\ \hline
 $2$ & $5.92$ & $67$ & $0.65$ & $-0.2148$ & $0.5723$ & $846$ & $865$ & $1004$\\
 $20$ & $5.44$ & $25$ & $0.54$ & $-0.1953$ & $0.2637$ & $379$ & $318$ & $363$
\end{tabular}
\caption{Summary of the fitting results to Classical Nucleation Theory.}
\label{tb:fitparams}
\end{table}
For $\epsilon=2$ we obtain an effective critical cluster size of $R^*=5.92$, a chemical
potential difference of $\Delta \mu=-0.2148$ and a surface tension of
$\gamma=0.5723$, and for $\epsilon=20$ we obtain $R^*=5.44$, $\Delta \mu=-0.1953$, and $\gamma=0.2637$. These values are comparable to previous work using the umbrella sampling technique~\cite{auer02}.

The last interface positions $\lambda_{n-1}$ and $\lambda_{n-2}$, which were placed automatically in the FFS simulations, coincide with the critical cluster size $\lambda^*$. This confirms that the interface placement in FFS simulations of ref.~\cite{kratzer13} can be used to estimate the location of the top of the barrier by setting a desired transition probability of $p_\text{des}=0.5$, which corresponds to the committor value at $\lambda^*$~\cite{bolhuis02}.

\section{Discussion}

We have presented simulation results on the liquid-fcc nucleation of charged \mamos~at two  phase points with different contact energies.

The formation of a crystal cluster is indicated by precursor clusters with local sixfold symmetry at an early stage. There are no correlations with the local density or fourfold symmetry in our simulations. The nucleation is therefore initiated by spontaneous local ordering into a bcc-like structure as predicted by Alexander and McTague~\cite{alexander78}, rather than density fluctuations, as previously discussed~\cite{russo12a,tan14}.

Our main observation is that the liquid-fcc nucleation of the \mamos~is a two-stage process passing through an intermediate bcc-like phase. Thus, there are two energy barriers towards the thermodynamically stable fcc-phase, which are dependent on the contact energy value. For a low contact energy, the second energy barrier can be overcome spontaneously and the transition can be treated as a direct liquid~$\rightarrow$~hcp/fcc-like transition, with only a weakly metastable bcc-like phase in between. In contrast, for a high contact energy, the system crystallizes with high probability into a bcc-like state. Starting from this phase, the second transformation to hcp/fcc occurs.
The second bcc-fcc transformation does not occur in a bulk bcc-like phase, but rather in a post-critical bcc-like cluster. This transformation again resembles a nucleation event, but this time of a cluster of fourfold local symmetry in a bcc crystallite. The transformation is greatly facilitated by the presence of the liquid-bcc surface, and thus, this second transformation is a heterogeneous nucleation, although we started from a homogeneous liquid. We quantified this process to occur with a transition rate of $9.52\times 10^{-08}\tau^{-1}\sigma^{-2}$ with respect to the surface of the critical-bcc cluster.

Despite of the fact that we observe a heterogeneous nucleation of the fcc phase from inside a bcc cluster, the free energy landscape is well-described by CNT. This is due to the fact that the crystallites are surprisingly spherical, and that the initial nucleation of the bcc crystallite is the rate limiting factor.

\section*{Acknowledgments}
A.A.~and K.K.~would like to thank the German Research Foundation (DFG) for financial support of the project within the Cluster of Excellence in Simulation Technology (EXC 310/1) at the University of Stuttgart. Computational resources were provided by the HLRS Stuttgart.

\appendix

\section{FFS details}\label{sec:appendix_FFS}

\subsection*{Simulation 1: $\epsilon=2$, $l=6$}
\begin{table}[H]
\centering
\begin{tabular}{c|ccccccc}
$i$ & 0 & 1 & 2 & 3 & 4 & 5 & 6\\ \hline
$\lambda_i$ & 5 & 21 & 44 & 69 & 92 & 116 & 145 \\ \vspace{0.3cm}
$p_i$ & -- & 0.002 & 0.002 & 0.004 & 0.016 & 0.011 & 0.009 \\
$i$ & 7 & 8 & 9 & 10 & 11 & 12 & 13 \\ \hline
$\lambda_i$ & 172 & 204 & 239 & 274 & 313 & 352 & 392 \\ \vspace{0.3cm}
$p_i$ & 0.021 & 0.017 & 0.021 & 0.026 & 0.037 & 0.058 & 0.068 \\
$i$ & 14 & 15 & 16 & 17 & 18 & 19 & 20 \\ \hline
$\lambda_i$ & 444 & 491 & 538 & 584 & 644 & 693 & 752 \\ \vspace{0.3cm}
$p_i$ & 0.051 & 0.119 & 0.143 & 0.213 & 0.202 & 0.375 & 0.401 \\
$i$ & 21 & 22 & 23 & 24 \\ \hline
$\lambda_i$ & 806 & 865 & 1004 & 7300 \\
$p_i$ & 0.519 & 0.675 & 0.634 & 0.901
\end{tabular}
\caption{Details of the FFS interfaces for $\epsilon=2$ when growing a cluster of solid-like particles with size $\lambda$.}
\end{table}

\subsection*{Simulation 2: $\epsilon=20$, $l=6$}
\begin{table}[H]
\centering
\begin{tabular}{c|ccccccc}
$i$ & 0 & 1 & 2 & 3 & 4 & 5 & 6 \\ \hline
$\lambda_i$ & 5 & 24 & 51 & 73 & 101 & 128 & 157 \\ \vspace{0.3cm}
$p_i$ & -- & 0.006 & 0.013 & 0.063 & 0.052 & 0.087 & 0.129 \\
$i$ & 7 & 8 & 9 & 10 & 11 & 12\\ \hline
$\lambda_i$ & 190 & 222 & 272 & 318 & 363 & 7300 \\
$p_i$ & 0.206 & 0.291 & 0.225 & 0.421 & 0.559 & 0.562
\end{tabular}
\caption{Details of the FFS interfaces for $\epsilon=20$ when growing a cluster of solid-like particles with size $\lambda$. }
\end{table}

\subsection*{Simulation 3: $\epsilon=20$, $l=4$}
\begin{table}[H]
\centering
\begin{tabular}{c|cccccc}
$i$ & 0 & 1 & 2 & 3 & 4 & 5 \\ \hline
$\lambda_i$ & 15 & 84 & 132 & 154 & 334 & 934  \\ \vspace{0.3cm}
$p_i$ & -- & 0.006 & 0.178 & 0.694 & 0.741 & 0.599  \\
$i$ & 6 & 7 & 8 & 9 & 10 & 11 \\ \hline
$\lambda_i$ & 1383 & 1852 & 2230 & 2805 & 3434 & 5200 \\
$p_i$ & 0.813 & 0.952 & 1.000 & 0.971 & 0.980 & 1.000
\end{tabular}
\caption{Details of the FFS interfaces for $\epsilon=20$ when growing an fcc-like cluster of size $\lambda$.}
\end{table}

\subsection*{Nucleation rates}
\begin{table}[H]
\begin{tabular}{c|c|c|c|c|c|c|c}
$\epsilon$ & $l$ & $\lambda_A$ & $\lambda_B$ & $n$ &$\Phi [\tau^{-1}\sigma^{-3}]$ & $P_B$ & $k_{AB} [\tau^{-1}\sigma^{-3}]$ \\ \hline
$2$ & $6$ & $5$ & $7300$ & $24$ & $9.15\times10^{-05}$ & $9.36\times10^{-31}$ & $8.54\times 10^{-35}$ \\
$20$ & $6$ & $5$ & $7300$ & $12$ & $7.27\times10^{-05}$ & $5.08\times 10^{-12}$ & $3.72\times 10^{-16}$ \\
\end{tabular}
\caption{FFS simulation results for the first two traces shown in Fig.~\ref{fig:real_cluster} which are the regular simulations driven by $\lambda_6$.
}
\label{tb:rates2_20}
\end{table}

\begin{table}[H]
\begin{tabular}{c|c|c|c|c|c|c|c}
$\epsilon$ & $l$ & $\lambda_A$ & $\lambda_B$ & $n$ &$\Phi [\tau^{-1}\sigma^{-2}]$ & $P_B$ & $k_{AB} [\tau^{-1}\sigma^{-2}]$ \\ \hline
$20$ & $4$ & $15$ & $5200$ & $11$ & $3.76\times10^{-04}$ & $2.58\times 10^{-04}$ & $9.52\times 10^{-08}$
\end{tabular}
\caption{FFS simulation results for the last trace shown in Fig.~\ref{fig:real_cluster}, optimizing for a larger fcc-like cluster with $\lambda_4$ as order parameter. Note, that the values are given with respect to the surface of the spherical cluster at the critical size (in $\sigma^{-2}$).
}
\label{tb:rates_fcc}
\end{table}

\end{document}